\documentclass[final,5p,times,twocolumn]{elsarticle}

\usepackage{amssymb}

\journal{Physics Letters B}

\biboptions{sort&compress}
\usepackage[all]{nowidow}
\usepackage{amsmath}
\newcommand*{\taukern}{\tau\mkern+1mu}

\begin{document}

\begin{frontmatter}



\title{Limiting fragmentation in heavy-ion stopping?}


\author{Johannes Hoelck$^1$, Emiko Hiyama$^2$, and Georg Wolschin$^{1,2}$}

\noindent
{
	{$^1$Institute for Theoretical Physics, Heidelberg University}\\
	\hspace{.6cm}
	{Philosophenweg 16}\\
	{Heidelberg~{69120}\\{Baden W\"urttemberg}, {Germany}}\\\\
	\hspace{.6cm}
	{$^2$Department of Physics, Tohoku University}\\
	{Sendai~{980-8578}, {Japan}} and\\
	{RIKEN Nishina centre, 2-1 Hirosawa,\\ Wako 351-0106, Japan}\\\\
}

\begin{abstract}
Based on a nonequilibrium--statistical relativistic diffusion model that is consistent with quantum chromodynamics (QCD), we investigate baryon stopping in relativistic heavy-ion collisions at SPS, RHIC, and LHC energies.
The net-proton rapidity distributions of the individual fragments exhibit a scaling behaviour similar to limiting fragmentation (LF) that is related to geometric scaling in the colour-glass condensate (CGC) and depends upon the gluon saturation scale.
Forward-angle net-proton data at energies reached at the LHC are required to verify the prediction.
\end{abstract}




\end{frontmatter}


\section{Introduction}
\label{intro}
Limiting fragmentation (LF) was first shown to occur in charged-hadron production at large pseudorapidities in the fragmentation region of $p\bar{p}$ data in an energy range of $\sqrt{s} = 53$--$900$\,GeV \cite{al86}.
Here, the charged-particle pseudorapidity yield $dN_\mathrm{ch}/d\eta$ does not depend on energy over a large range of pseudorapidities $\tilde{\eta}=\eta - y_\mathrm{beam}$, with the beam rapidity $y_\mathrm{beam}$.
The phenomenon had been predicted earlier theoretically for hadron--hadron and electron--proton collisions \cite{ben69}.

In the context of relativistic heavy-ion physics, it was first shown at the Relativistic Heavy Ion Collider (RHIC) in 2002 that the approach to a universal limiting curve is a characteristic feature of the particle production process, and this was eventually confirmed in the energy range $\sqrt{s_\mathrm{NN}} = 19.6$--$200$\,GeV \cite{bea02,bb03,ada06}.
Here, the LF hypothesis also holds in a given centrality class within the experimental error bars.

Since forward-rapidity data are missing at the Large Hadron Collider (LHC), it is presently unclear whether LF in hadron production is fulfilled at energies reached at the LHC, with the emphasis on $2.76$\,TeV and -- in the forthcoming Run 3 -- $5.36$\,TeV Pb--Pb collisions.
Various model calculations have come to differing conclusions \cite{sta06,sahoo19,tor19,kgw19}, with our phenomenological three-source relativistic diffusion model (RDM) predicting that LF in charged-hadron production should be approximately fulfilled at LHC energies as well \cite{kgw19,kgw21}, even though the cross sections rise with increasing energy.

Apart from the LF behaviour of produced charged hadrons in relativistic heavy-ion collisions, it is of interest to investigate the scaling properties of the stopping distributions of net baryons.
These are much more sensitive to the initial-state physics, because stopping occurs on a very short timescale of $t < 0.1$\,fm/$c$ that is of the order of the local thermalization time for gluons \cite{gw22}.
The stopping process can be measured through the net-proton distributions (protons minus produced antiprotons), with some uncertainty in the conversion to net baryons \cite{app99}, which are the conserved quantity.

In this Letter, we account for the time-dependence of the stopping process and the accompanying rapidity loss of the net-baryon distributions in a relativistic diffusion model that is consistent with QCD, and predict the LF properties of the net-baryon distributions from SPS to LHC energies.
In Section 2, an outline of the time-dependent model \cite{hgw20} is given.
In Section 3, we focus on the time-asymptotic equilibrium solutions that coincide with the QCD-inspired stopping model of Refs.\,\cite{mtw09,mtwc09}, and investigate their scaling behaviour, comparing to data from SPS and RHIC.
The linear dependence of the position of the stopping peak on the beam rapidity is discussed.
The conclusions are drawn in Section 4.


\section{A nonequilibrium--statistical stopping model}
\label{nonequilibrium}
Baryon stopping is modelled as a diffusive process in rapidity space.
Our approach is inspired by the phenomenological relativistic diffusion model \cite{gw99} and uses similar key assumptions, but is based on stochastic particle trajectories constructed from relativistic Markov processes in phase space \cite{hgw20}.
The latter reduce to non-Markovian stochastic processes in position space, consistent with the requirements of special relativity \cite{lopuszanski-1953}.
Time evolution is governed by a fluctuating background that represents the partons of the fragments.
The nucleon distribution function that is shaped by the interactions of the valence quarks with the fluctuating background -- in particular, soft gluons in the other nucleus -- can then be expressed as a superposition of time-dependent single-particle probability density functions.
To determine the coefficient functions of the associated drift--diffusion processes, fluctuation--dissipation relations are derived from the particles' expected time-asymptotic behaviour, which allows us to construct transport coefficients for stopping that are physically motivated from QCD.

Starting at time $t = t_\mathrm{i}$ from a given initial distribution that is provided by the Fermi-gas model, we take any particle trajectory to evolve in time towards an expected asymptotic equilibrium state.
Here, the term `equilibrium' refers to a stationary state
that is \emph{not} thermal; it is reminiscent of a nonthermal fixed point.
Since strong interactions in relativistic heavy-ion collisions effectively cease at a finite interaction time, the time evolution is terminated at $t = t_\mathrm{f} > t_\mathrm{i}$ before the equilibrium state is reached, so that the system remains in a final nonequilibrium state.

\bigbreak

In this work, we concentrate on the rapidity variable for net baryons, or net protons, to investigate the scaling behaviour in stopping.
Because net-baryon distributions cannot be accessed experimentally, we either consider participant protons, and compare to measured net-proton number density distributions in rapidity space, or net-baryon distributions whenever these have been constructed by the experimental collaborations from their net-proton data.

We incorporate the spatial separation of the two nuclear fragments through a two-source ansatz~\cite{gw99}.
The time evolution of particles originating from the forward- and backward-going fragments is then represented through separate probability densities and fluctuation--dissipation relations.
Since the system is symmetric with respect to its centre of momentum, the net-proton number density in rapidity space in the system's centre-of-momentum frame can be expressed in terms of the superposition of the forward- and backward-going distributions as
\begin{equation}
	\label{dNdy}
	\frac{dN_{p-\bar{p}}}{dy}(t;y) \simeq \frac{N_{p-\bar{p}}}{2} \left[\psi(t;+y) + \psi(t;-y)\right]
	\mathinner{.}
\end{equation}
Here, ${N_{p-\bar{p}}}$ is the net-proton number, and $\psi(t;y)\,dy$ the probability to find a participant proton from the forward-going fragment at time~$t$ with rapidity in $[y,y+dy]$.

\begin{figure}[t]
	\centering%
	\includegraphics[scale=1.1]{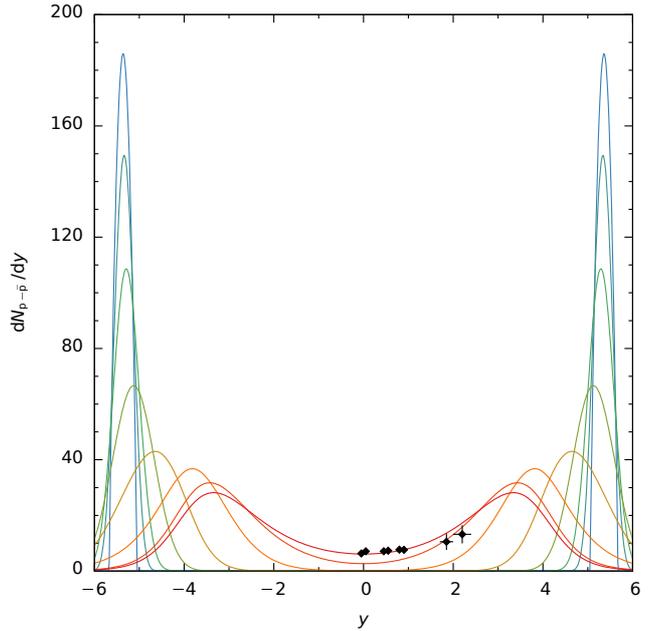}%
	\caption{
		Nonequilibrium net-proton rapidity distribution functions for central Au--Au collisions with centre-of-mass energy $\sqrt{s_\mathrm{NN}} = 200$\,GeV at {$0$--$5\%$} centrality as calculated in the relativistic diffusion model \cite{hgw20}.
		Solid lines mark different interaction times $(t_\mathrm{f} - t_\mathrm{i}) \times D = {0; 0.01; 0.03; 0.1; 0.3; 1; 3; 10}$.
		The latter are compared with experimental data (black circles) recorded earlier at RHIC by the BRAHMS Collaboration~\cite{bea04}.
	}%
	\label{fig1}%
\end{figure}

As initial state in the time evolution, $\psi_\mathrm{i}(y) \equiv \psi(t_\mathrm{i};y)$, we approximate each nucleus by a zero-temperature gas with Fermi momentum~$p_\mathrm{F}$, corresponding to the Fermi rapidity $y_\mathrm{F} = \operatorname{asinh}(p_\mathrm{F}/m)$.
Following the steps outlined in Ref.\,\cite{hgw20}, this leads to an initial distribution in rapidity space (outer peaked blue curves in Fig.\,\ref{fig1} for central $\sqrt{s_\mathrm{NN}}=200$\,GeV Au--Au, centre-of-momentum frame)
\begin{multline}
	\label{ini}
	\psi_\mathrm{i}(y_* + y_\mathrm{beam}) =
	\\
	\frac{1}{2} \sinh(y_\mathrm{F})^{-3} \, \Theta(y_\mathrm{F} - |y_*|) \, \cosh(y_*) \left[\left( {\frac{\cosh(y_\mathrm{F})}{\cosh(y_*)}}\right)^3 - 1\right]
	\mathinner{,}
\end{multline}
where $y_*$ denotes the proton rapidity in the rest frame of the forward-going nucleus.

Based on a description of the baryon trajectories as relativistic Markov processes in phase space, we obtain in Ref.\,\cite{hgw20} a Kramers equation for the marginal probability density function $f(t;x^3,y)$ of longitudinal position~$x^3$ and rapidity~$y$.
Since $x^3$ is unobservable, we integrate it out, resulting in a Fokker--Planck equation (FPE) for the marginal probability density function~$\psi(t;y)$
\begin{gather}
	\label{fpe}
	\partial_t \psi(t;y) = -\partial_y \left[\mu(y) \, \psi(t;y)\right] + D \, \partial_y^2 \psi(t;y)
	\mathinner{,}\\
	\psi(t;y) = \int d{x^3} \, f(t;x^3,y)
	\mathinner{.}
\end{gather}

The drift function~$\mu(y)$ and the diffusion coefficient~$D$ -- which we assume to be constant with respect to rapidity at a given energy in the present work -- are derived from the expected mesoscopic behaviour as proposed in Refs.\,\cite{deb97,du09}, rather than from microscopic considerations, which are more difficult to assess.
For this, a fluctuation--dissipation relation between drift and diffusion coefficient is established by differentiating the stationary equilibrium solution~$\psi_\mathrm{eq}$ of the FPE ($\partial_t \psi_\mathrm{eq} = 0$) with respect to rapidity,
\begin{equation}
	\label{fdr}
	\frac{\mu(y)}{D} = \partial_y \ln\bigl[\psi_\mathrm{eq}(y)\bigr]
	\mathinner{,}
\end{equation}
such that the drift becomes a function of $D$ and $y$.
Here, we identify the equilibrium~$\psi_\mathrm{eq}$ with a state created by inelastic scattering off a color-glass condensate (CGC).
More details on the underlying formalism are given in Section~\ref{equilibrium}.

For comparisons with experimental data, the solutions of the FPE
are evaluated at the final time~$t_\mathrm{f}$, when partonic interactions between the receding nuclei cease due to their increasing distance.
This time is, however, not an observable quantity in heavy-ion collisions, and in the actual solutions it appears always in products with other quantities, such as the diffusion coefficients (see below).
The latter can be determined in $\chi^2$ minimizations to the data.

With the drift function~$\mu(y)$, the diffusion constant~$D$, and the initial distribution Eq.\,(\ref{ini}), the FPE can be written in dimensionless form by substituting the time~$t$ with the dimensionless evolution parameter $\delta = (t - t_\mathrm{i}) / (t_\mathrm{f} - t_\mathrm{i})$.
It is solved numerically for $0 < \delta \le 1$ as detailed in Ref.\,\cite{hgw20}, where we had already presented results for central Pb--Pb at $\sqrt{s_\mathrm{NN}}=17.3$\,GeV and Au--Au at $\sqrt{s_\mathrm{NN}}=62.4$\,GeV.
Here, we compute corresponding results for central $200$\,GeV Au--Au, with several interaction times $t_\mathrm{f} - t_\mathrm{i}$ shown in Fig.\,\ref{fig1} in comparison with the $0$--$5\%$ BRAHMS stopping data \cite{bea04}.

When a linear approximation for the drift function is used together with a constant diffusion coefficient and a simplified initial state, analytical solutions of the FPE can directly be compared to data \cite{gw99,gw16}.

\section{Equilibrium solutions and scaling in stopping}
\label{equilibrium}
If the stochastic process that accounts for stopping
continued past $t = t_\mathrm{f}$, it would converge to a stationary equilibrium state.
We take this state to arise from the inelastic scattering of the protons' valence quarks with a colour-glass condensate (CGC) \cite{gri83,mue86,jpbmue87,mcl94}, a coherent state based on the saturation of the gluon density below a characteristic momentum scale~$Q_{\mathrm{s}}$.
The CGC framework is an effective theory for high density matter of saturated gluons that is based on QCD, which is expected to hold in the present context.
An important property of cross sections derived in the CGC framework is geometric scaling: The energy dependence is fully determined through $Q^2/Q_\mathrm{s}^2 \equiv \zeta$, as used, e.g., in the 
models of Ref.\,\cite{gbw98} for deep-inelastic electron--proton scattering, or Ref.\,\cite{mtw09} for heavy-ion collisions, where the momentum exchange is approximated by the transverse momentum of the produced hadron, $Q^2 \simeq (p^1)^2 + (p^2)^2$.
Since we incorporate this scaling property in the propagation of the time-dependent distribution functions via $\mu(y)$, our approach is consistent with the CGC framework, and thereby with QCD.

The equilibrium distribution of the forward-going (`projectile-like') participant protons is then given by \cite{kha04,bai06,dum06}
\begin{equation}
	\label{pdf-dis}
	\psi_\mathrm{eq}(y) = \frac{C}{2\pi} \int_0^1 d{x} \, {q_v(x)} \, {g(x^{2+\lambda} \, e^{\taukern(y)})}
	\mathinner{.}
\end{equation}
The longitudinal momentum fraction carried by the protons' valence quarks is $x$, and $q_v$ denotes the valence-quark distribution function calculated from the NNLO result \cite{mrst02}.
The normalizing constant~$C$ sets the integral of $\psi_\mathrm{eq}$ to unity. We solve the integral in Eq.\,(\ref{pdf-dis}) numerically with adaptive Gauss--Kronrod quadrature in the full rapidity space.
For sufficiently large rapidities, also analytical approximate solutions exist (see below), which can be used to check the accuracy of the full numerical solutions in the tails.

The distribution function~$g$ of the condensed soft gluons in the backward-going fragment can be reduced to a simple function of the scaling variable $\zeta = \left[(p^1)^2 + (p^2)^2\right] / {Q_{\mathrm{s}}^2}$ using the Golec-Biernat--W{\"u}sthoff model \cite{gbw98},
\begin{equation}
	g(\zeta) = 4\pi \zeta e^{-\zeta}
	\mathinner{.}
	\label{gbw}
\end{equation}
The $x$ dependence of $Q_{\mathrm{s}}$ is determined by the gluon-saturation-scale exponent~$\lambda$
\begin{equation}
	Q_{\mathrm{s}}^2 =Q_0^2 \, A^{1/3} \, x^{-\lambda}
	\mathinner{,}
\end{equation}
where the constant~$Q_0^2$ sets the dimension, and the mass number~$A$ the scaling with the nuclear size.
For a given centre-of-mass energy per nucleon pair~$\sqrt{s_\mathrm{NN}}$, these three parameters completely determine the rapidity dependence of $\psi_\mathrm{eq}$ through the dimensionless function
\begin{equation}
	\label{tau}
	\begin{split}
		\taukern(y)
		&= \ln\left(\frac{s_\mathrm{NN}}{Q_0^2}\right) - \frac{1}{3} \ln(A) - 2 \, (1 + \lambda) \, y
		\\
		&= 2 y_\mathrm{beam} + \ln\left(\frac{m_p^2}{Q_0^2}\right) - \frac{1}{3} \ln(A) - 2 \, (1 + \lambda) \, y
		\mathinner{.}
	\end{split}
\end{equation}

This formalism had already been used in Refs.\,\cite{mtw09,mtwc09}, where we had fitted corresponding distribution functions directly to stopping data at SPS and RHIC energies, without considering a time evolution of the system.
Typical values for the saturation-scale exponent in stopping are $\lambda=0.2$--$0.3$, and we use $\lambda=0.2$ together with $Q_0^2=0.09$\,GeV$^2$ in this work.
These values compare well with fit results from deep-inelastic electron--proton data from the DESY Hadron--Electron Ring Accelerator HERA, where $\lambda\simeq 0.288$ and $Q_0^2\simeq 0.097$\,GeV$^2$ \cite{gbw98}.

\begin{figure}[t]
	\centering%
	\includegraphics[scale=1.1]{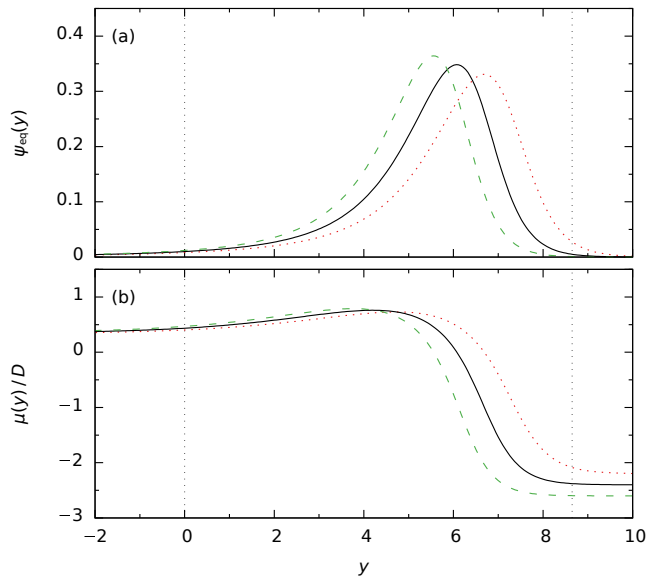}%
	\caption{
		Stationary distribution function (a) and fluctuation--dissipation relation (b) for $\psi_\mathrm{eq}$ of the forward-moving nucleus in a central collision of Pb nuclei with centre-of-mass energy $\sqrt{s_\mathrm{NN}}=5.36$\,TeV for three different values of the gluon-saturation-scale exponent:
		$\lambda=0.1$ (dotted), $0.2$ (solid), and $0.3$ (dashed). The right dotted vertical line indicates the beam rapidity $y_\mathrm{beam}=8.651$.
		Stopping increases with rising $\lambda$.
	}%
	\label{fig2}%
\end{figure}

The amount of stopping depends on the gluon saturation momentum~$Q_\mathrm{s}$, and hence, on the saturation-scale exponent~$\lambda$.
In Fig.\,\ref{fig2}, we show the stationary forward-going distribution~$\psi_\mathrm{eq}$ from Eq.\,(\ref{pdf-dis}) in central $5.36$\,TeV Pb--Pb collisions together with the corresponding fluctuation--dissipation relation Eq.\,(\ref{fdr}) for $Q_0^2=0.09$\,GeV$^2$ and three different values of the saturation-scale exponent $\lambda=0.1$ (dotted), $0.2$ (solid), and $0.3$ (dashed), or $Q_\mathrm{s}\simeq 1.03$, $1.46$, and $2.06$\,GeV at $x=10^{-3}$.
Stopping is seen to increase with an increasing gluon saturation scale,
and thereby affects the longitudinal scaling behaviour. For $\lambda>0$, when the gluon saturation scale depends on the gluon's Bjorken-$x$, limiting fragmentation is broken in our model, and the size of the LF violation increases with $\lambda$.

As detailed in the full account of our nonequilibrium--statistical model in Ref.\,\cite{hgw20},
$\psi_\mathrm{eq}(y)$ in Eq.\,(\ref{pdf-dis}) decays
exponentially at large positive and negative values of the rapidity.
In the backward-going region, in particular, the exponential damping with $\taukern(y)$ causes only small $x$ values to contribute.
Here, the valence-quark distribution is $x \mathinner{q_v(x)} \sim a x^b$, and Eq.\,(\ref{pdf-dis}) becomes 
\begin{equation}
	\psi_\mathrm{eq}(y) \underset{y \to -\infty}{\sim} \exp(\alpha_- y + \mathrm{const.})
\end{equation}
with $\alpha_- = 2 b \, (1 + \lambda) / (2 + \lambda)$.
The corresponding decay rate at large positive rapidities is $\alpha_+= -2 \, (1 + \lambda)$.
As a consequence,
the drift function~$\mu(y)$ defined via Eq.\,(\ref{fdr}) does not depend on rapidity for $y\rightarrow \pm \infty$ and constant diffusion~$D$.



Whereas the time dependence of the stopping process
as discussed in the previous section is itself of physical interest, it is eventually determined by the equilibrium distributions whether limiting fragmentation (LF) or a comparable scaling behaviour is present in stopping, or not.
In Ref.\,\cite{hgw20}, we found the final nonequilibrium distributions to be close to equilibrium at high RHIC and LHC energies, and it is therefore sufficient to study the scaling behaviour of the equilibrium state, where analytic calculations are possible \cite{mtw09}.

\begin{figure}[t]
	\centering%
	\includegraphics[scale=1.1]{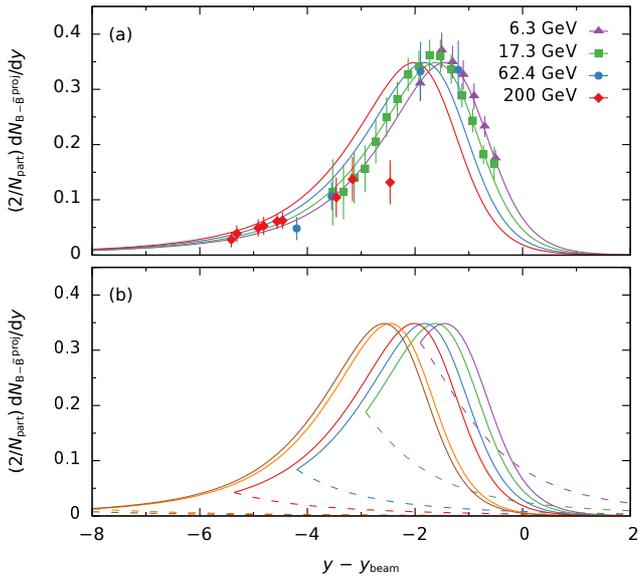}%
	\caption{
		(a) Experimental normalized net-baryon rapidity distributions as functions of $y-y_\mathrm{beam}$ for central Pb--Pb at $\sqrt{s_\mathrm{NN}} = 6.3$\,GeV \cite{blu07} (triangles) and $17.3$\,GeV \cite{app99} (squares), central Au--Au at $\sqrt{s_\mathrm{NN}} = 62.4$\,GeV \cite{ars09} (circles) and $200$\,GeV \cite{bea04} (diamonds).
		Contributions from the backward-going sources have been subtracted, see text.
		LF scaling is almost fulfilled at SPS and RHIC energies within the error bars, but LHC data are not available.
		Solid curves are our results for the forward-going fragments.
		(b) Calculated equilibrium net-baryon distribution functions at $\sqrt{s_\mathrm{NN}} = 6.3$, $17.3$, $62.4$, $200$, $2760$, and $5362$\,GeV (right to left).
		Contributions from the forward-going sources are solid, from the backward-going sources dashed.
		The forward contributions from the backward-going sources are negligible at both LHC-energies.
		In the model, LF scaling in the variable $y-y_\mathrm{beam}$ is violated in stopping.
	}%
	\label{fig3}%
\end{figure}
\begin{figure}[t]
	\centering%
	\includegraphics[scale=1.1]{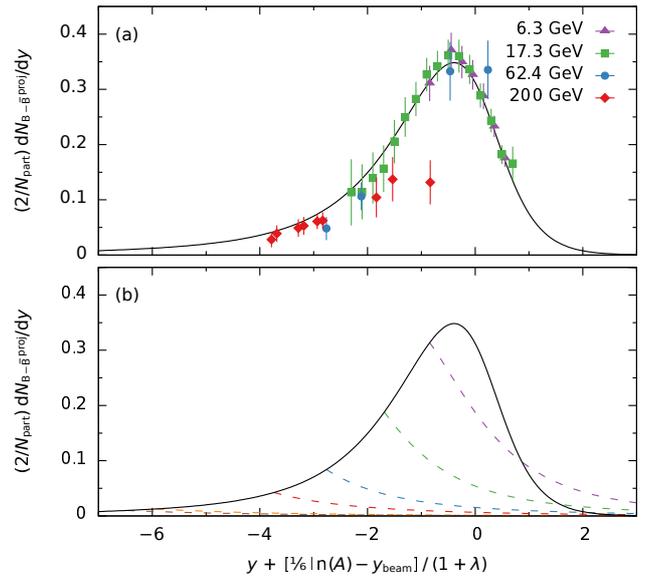}%
	\caption{
		(a) Experimental normalized net-baryon rapidity distributions as functions of $y+[\ln(A)/6-y_\mathrm{beam}]/(1+\lambda)$ for central Pb--Pb at $\sqrt{s_\mathrm{NN}} = 6.3$\,GeV \cite{blu07} (triangles) and $17.3$\,GeV \cite{app99} (squares), central Au--Au at $\sqrt{s_\mathrm{NN}} = 62.4$\,GeV \cite{ars09} (circles) and $200$\,GeV \cite{bea04} (diamonds).
		Contributions from the backward-going sources have been subtracted.
		The solid cuve is the corresponding model result.
		LF scaling is fulfilled within the error bars.
		(b) Calculated equilibrium net-baryon distributions for central Pb--Pb and Au--Au at $\sqrt{s_\mathrm{NN}} = 6.3$, $17.3$, $62.4$, $200$, $2760$, and $5362$\,GeV (top to bottom) as functions of the variable $y+[\ln(A)/6-y_\mathrm{beam}]/(1+\lambda)$.
		Contributions from the forward-going sources are solid, from the backward-going sources dashed.
		The model results -- here, for $\lambda=0.2$ and $Q_0^2=0.09$\,GeV$^2$ -- exhibit exact scaling as function of this variable.
	}%
	\label{fig4}%
\end{figure}
\begin{figure}[t]
	\centering%
	\includegraphics[scale=0.28]{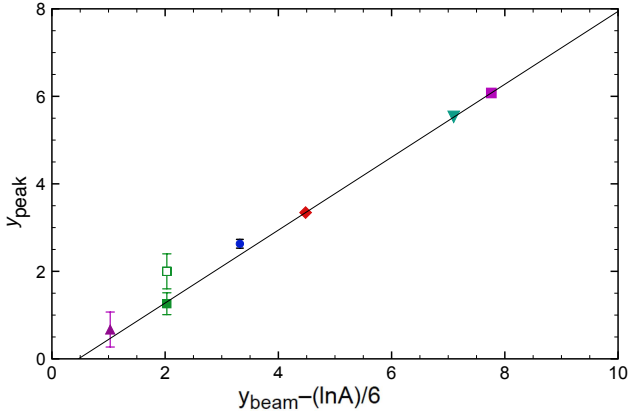}%
	\caption{
		Calculated positions~$y_\mathrm{peak}$ of the stopping peaks as function of the beam rapidity~$y_\mathrm{beam}$ (solid line) compared to corresponding peak positions as inferred \cite{mtw11} from data for central Pb--Pb at $\sqrt{s_\mathrm{NN}} = 6.3$\,GeV \cite{blu07} (triangle), $17.3$\,GeV \cite{app99,blu07} (filled and open squares), and central Au--Au at $\sqrt{s_\mathrm{NN}} = 62.4$\,GeV \cite{ars09} (circle).
		The stopping-peak positions for $200$\,GeV Au--Au (diamond), $2.76$\,TeV Pb--Pb (inverted triangle), and $5.36$\,TeV Pb--Pb (square) are as calculated in the model.
		The gluon-saturation-scale parameters are $\lambda=0.2$ and $Q_0^2 = 0.09$\,GeV$^2$.
	}%
	\label{fig5}%
\end{figure}

To properly account for the scaling behaviour, we deduct from the data the respective backward-going contributions that we calculate in our model in order to isolate the net-baryon rapidity distribution for the forward-going source.
Contributions from the backward-going source are small at LHC energies, but significant at RHIC energies, and even more so at SPS energies where the separation of the stopping peaks in rapidity space can become so small that they appear as a single peak.

Considering only the forward-going distributions at six centre-of-mass energies as functions of the LF variable $y-y_\mathrm{beam}$, we compare in Fig.\,\ref{fig3} with the corresponding SPS and RHIC data.
The deducted contributions from the backward-going sources are shown as dashed curves in the lower frame.
Apart from the outlying point\footnote{this point was omitted in Fig.1} at $200$\,GeV, limiting fragmentation seems to be almost fulfilled at SPS and RHIC energies within the experimental error bars, but at $6.3$\,GeV and $17.3$\,GeV, the small violation of LF scaling that our model predicts actually agrees with the data.

The solid curves are the normalized stopping distributions for the forward-going (`projectile') source. These are repeated in the lower frame together with our results at LHC energies.
LF scaling as function of the variable $y-y_\mathrm{beam}$ is clearly not fulfilled in the model calculations, although the differences are small at SPS and RHIC energies as shown in the upper frame. Indeed, LF has recently assumed to be valid in baryon stopping by Braun-Munzinger et al.~\cite{pbm21}.
At present, however, only at SPS energies significant data beyond the stopping peak exist.

It is interesting that the model is compatible with a slightly different scaling relation, which is based on geometric scaling:
Since the function~$\taukern(y)$ from Eq.\,(\ref{tau}) encodes the entire $y$ dependence of the equilibrium distribution~$\psi_\mathrm{eq}(y)$, the latter becomes independent of $\sqrt{s_\mathrm{NN}}$, $A$, $\lambda$, and $Q_0$ when plotted against $\taukern(y)$.
Assuming that the saturation-scale parameters $\lambda$ and $Q_0$ agree for all collisions under consideration, this is also true for
\begin{equation}
	\frac{\ln\left(\frac{m_p^2}{Q_0^2}\right) - \taukern(y)}{2 \, (1 + \lambda)} = y + \frac{\frac{1}{6} \ln(A) - y_\mathrm{beam}}{1 + \lambda}
	\mathinner{,}
\end{equation}
as shown in Fig.\,\ref{fig4}.
A corresponding invariance has already been observed in our prior analysis \cite{mtwc09}.

The above is especially fulfilled at the position of the stopping peak, $y = y_\mathrm{peak}$, which establishes a linear relation between $y_\mathrm{peak}$ and the beam rapidity~$y_\mathrm{beam}$,
\begin{equation}
	y_\mathrm{peak} = c_0 + c_1 \, \bigl[y_\mathrm{beam} - \tfrac{1}{6} \ln(A)\bigr]
	\mathinner{,}
\end{equation}
with the energy- and nucleus-independent coefficients
\begin{align}
	\label{peak}
	c_0 &= \frac{\ln\left(\frac{m_p}{Q_0}\right) - \frac{1}{2} \taukern(y_\mathrm{peak})}{1 + \lambda}
	\mathinner{,}&
	c_1 &= \frac{1}{1 + \lambda}
	\mathinner{.}
\end{align}
Here, $\taukern(y_\mathrm{peak})$ implicitly depends on $\lambda$ and can be obtained by maximizing Eq.\,(\ref{pdf-dis}).
This relation is seen to be rather well fulfilled at SPS and RHIC in Fig.\,\ref{fig5} for $\lambda = 0.2$, $Q_0^2 = 0.09$\,GeV$^2$ ($c_0 \simeq -0.39$, $c_1 = 0.8\overline{3}$), although the $17.3$\,GeV results are somewhat uncertain because two successive measurements at SPS \cite{app99,blu07} differ slightly, and significant data beyond the peaks are not available at RHIC energies.
It would therefore be valuable to obtain LHC data for the stopping peak positions, as indicated in the figure, upper two symbols.

\section{Conclusions}
We have introduced a nonequilibrium--statistical diffusion model for stopping in relativistic heavy-ion collisions that accounts for the time-dependence of the stopping process, and is consistent with QCD.
While experimental data suggest that limiting fragmentation (LF) in the variable $y-y_\mathrm{beam}$ is fulfilled within the error bars for net-baryon rapidity distributions from SPS to RHIC energies, our model yields in accordance with Ref.\,\cite{mtw09} a slightly different scaling behaviour in central collisions up to LHC energies
that arises from geometric scaling in the CGC. For noncentral collisions, similar results can be expected.

This is in contrast to charged-hadron production, where not only data from SPS to RHIC energies agree with LF, but also our phenomenological model calculations \cite{kgw19,kgw21} predict that LF is approximately fulfilled in the variable $\eta-y_\mathrm{beam}$ from SPS to LHC energies.
The reason for this qualitative difference in the scaling properties of stopping and particle production could be the very different kinematics in the time evolution of incoming baryons, as compared to produced charged hadrons.
In general, the stopping distributions that are generated on a very short timescale are much more sensitive to the initial hard interactions than the (pseudo-)rapidity distributions of light charged hadrons that are mostly produced at the phase boundary, and hence, at a much later stage.
Moreover, in particle production, the (unmeasurable) fragmentation peaks
are much closer to midrapidity than in baryon stopping, where they reflect the kinematics of the incoming baryons more directly.

The degree of LF violation in baryon stopping as predicted by our model depends on the centre-of-mass energy, the gluon saturation scale, and to a lesser extent on the mass of the colliding nuclei.
To verify this conclusion, forward-rapidity net-proton data for Pb--Pb collisions at LHC energies would be needed.
Since their measurement using central Pb--Pb collisions in the LHCb detector will likely not be possible, this would require a new forward spectrometer in ALICE that is capable of measuring identified protons at very small angles, corresponding to rapidities $y \simeq 6$--$8$.

\section*{Acknowledgements}
Discussions with Peter Braun-Munzinger and Johanna Stachel are gratefully appreciated.
One of the authors (GW) acknowledges the support of the Japan Society for the Promotion of Science (JSPS) through BRIDGE fellowship BR200102 at Tohoku University (Sendai), RIKEN (Wako) -- where most of this manuscript has been written --, and the University of Tokyo.









\IfFileExists{gw_23.bib}{
	\bibliographystyle{elsarticle-num}
	\bibliography{gw_23}
}{

}

\end{document}